\include{BoxedEPS}
\include{BoxedEPS}
\documentclass[12pt]{amsart}
\usepackage{graphicx,amssymb,amsfonts,epsfig,amsthm,a4,amsmath,url,color,amsbsy,mathrsfs}
\usepackage{pdfsync}
\input amssym.def
\input amssym
\chardef\bslash=`\\ % p. 424, TeXbook
\makeatletter
\def\verbatim{\interlinepenalty\@M \@verbatim
   \leftskip\@totalleftmargin\advance\leftskip2pc
   \frenchspacing\@vobeyspaces \@xverbatim}
\makeatother
\newtheorem{thm}{Theorem}[section]
\newtheorem{cor}[thm]{Corollary}
\newtheorem{lem}[thm]{Lemma}
\newtheorem{prop}[thm]{Proposition}

\theoremstyle{definition}

\theoremstyle{remark}
\newtheorem{rem}{Remark}[section]
\newtheorem{exmp}{Example}[section]

\numberwithin{equation}{section}

\newcommand{\begeq}{\begin {equation}}
\newcommand{\eq}{\end{equation}}
\newcommand{\bs}{\begin {split}}
\newcommand{\es}{\end{split}}
\newcommand{\bp}{\begin {prop}}
\newcommand{\ep}{\end {prop}}
\newcommand{\bt}{\begin {thm}}
\newcommand{\et}{\end {thm}}
\newcommand{\bc}{\begin {cor}}
\newcommand{\ec}{\end {cor}}
\newcommand{\bl}{\begin {lem}}
\newcommand{\el}{\end {lem}}
\newcommand{\bpf}{\begin {proof}}
\newcommand{\epf}{\end {proof}}
\newcommand{\bi}{\begin {itemize}}
\newcommand{\ei}{\end {itemize}}
\newcommand{\ben}{\begin {enumerate}}
\newcommand{\een}{\end {enumerate}}
\newcommand{\brem}{\begin {rem}}
\newcommand{\erem}{\end {rem}}
\newcommand{\bex}{\begin {exmp}}
\newcommand{\eex}{\end {exmp}}

\newcommand{\cA}{\mathcal A}

\newcommand{\ZZ}{{\mathbb Z}}
\newcommand{\TT}{{\mathbb T}}
\newcommand{\RR}{{\mathbb R}}
\newcommand{\CC}{{\mathbb C}}
\newcommand{\NN}{{\mathbb N}}

\newcommand{\A}{{A}_\Omega}
\newcommand{\vecx}{\bar{\mathbf{x} }}
\newcommand{\vecy}{\bar{\mathbf{y} }}

\newcommand{\vecv}{\bar{\mathbf{v} }}
\newcommand{\vecu}{\bar{\mathbf{u} }}
\newcommand{\vecw}{\bar{\mathbf{w} }}

\DeclareMathOperator{\sinc}{sinc}

\DeclareMathOperator{\ssup}{ess\, sup}

\DeclareMathOperator{\dist}{dist }

\begin{document}

\title[ ] {Exact reconstruction of spatially undersampled signals in evolutionary systems}
\author{A. Aldroubi, J. Davis, and I. Krishtal}
\address{Department of Mathematics, Vanderbilt University, Nashville, TN 37240 \\
email: aldroubi@math.vanderbilt.edu, jacquedavis@gmail.com}
\address{Department of Mathematical Sciences, Northern Illinois University 
Watson Hall 320,
DeKalb, IL 60115 \\ email: krishtal@math.niu.edu}

\thanks{Partially supported by NSF grant  }

\maketitle

\begin{abstract}
 We consider the problem of spatiotemporal sampling in which an initial state $f$  of an evolution process $f_t=A_tf$ is to be recovered from a combined set of coarse samples from varying time levels $\{t_1,\dots,t_N\}$. This new way of sampling, which we call dynamical sampling, differs from standard sampling since at any fixed time $t_i$ there are not enough samples to recover the function $f$  or the state $f_{t_i}$. Although dynamical sampling is an inverse problem, it  differs from the typical  inverse problems in which $f$ is to be recovered from $A_Tf$ for a single time $T$.  
 In this paper, we consider signals that are modeled by $\ell^2(\ZZ)$ or a shift invariant space $V\subset L^2(\RR)$.
\end{abstract}

\section{Introduction.}
In sampling theory we seek to reconstruct a function $f$ from its samples $\{f(x_j): \; x_j\in X\}$ where $X \subset \RR$ is a countable set. 
Perhaps the most well-known result  is the Shannon Sampling Theorem \cite{S98}. Specifically, if a function $f\in L^2(\RR)$ is $T$-bandlimited, i.e.,~its Fourier transform 
\[\hat f(\xi) = \int_{\RR} f(x)e^{-2\pi i x\xi}dx, \ \xi\in\RR,\]
has support contained in $[-T,T]$, then
\begeq
f(x) = \sum_{n=-\infty}^\infty f(\frac{n}{2T})\frac{\sin\pi(2Tx - n)}{\pi(2Tx - n)} =
\sum_{n=-\infty}^\infty f(\frac{n}{2T}){\sinc(2Tx - n)},\ x\in\RR,
\eq
where the series converges in  $L^2(\RR)$ and uniformly on compact sets. Thus,  in this situation, $f$ can be recovered from its samples $\{f(\frac{n}{2T}):\, n \in \ZZ\}$.

However, there are many situations in which the sampling of a function $f$ is restricted. For example,  assume that we wish to find a function $f$ in $[0,1]$ at time $t=0$ at a spatial resolution of $0.1$, i.e., we need to know the values of $f$ on the set $\{0.1k:\; k=1,\dots, 10\}$. Practical considerations, however, dictate that we can only use two sampling devices, i.e., we can only sample at two locations $\{x_1,x_2\}\subset \{0.1k:\; k=0,\dots, 10\}$. Can we still recover  $f$  at the spatial  resolution of $0.1$? The answer is that it may be possible  to determine $f$ at the correct resolution if we know that $f$ is evolving in time under the action of a known operator, such as  diffusion. This toy problem illustrates  a new type of sampling problems that we call dynamical sampling.
\vspace {2mm}

\noindent{\bf The dynamical sampling problem}. Assume that a function $f$ on a domain $D$ is an initial state of a physical process evolving in time under the action of a family of operators $A_t$ indexed by $t\ge 0$. Can we recover $f$ from the samples $\{f(X),f_{t_1}(X), \dots, f_{t_N}(X)\}$ of $f$ on $X\subset D$ and its various states  $f_t(X):=(A_tf)(X)$ at times $\{t_1,\dots,t_N\}$?
\vspace {2mm}

\begin{figure}\label{Fig1}
  \begin{center}
  \label {SimpleEx}
    \includegraphics{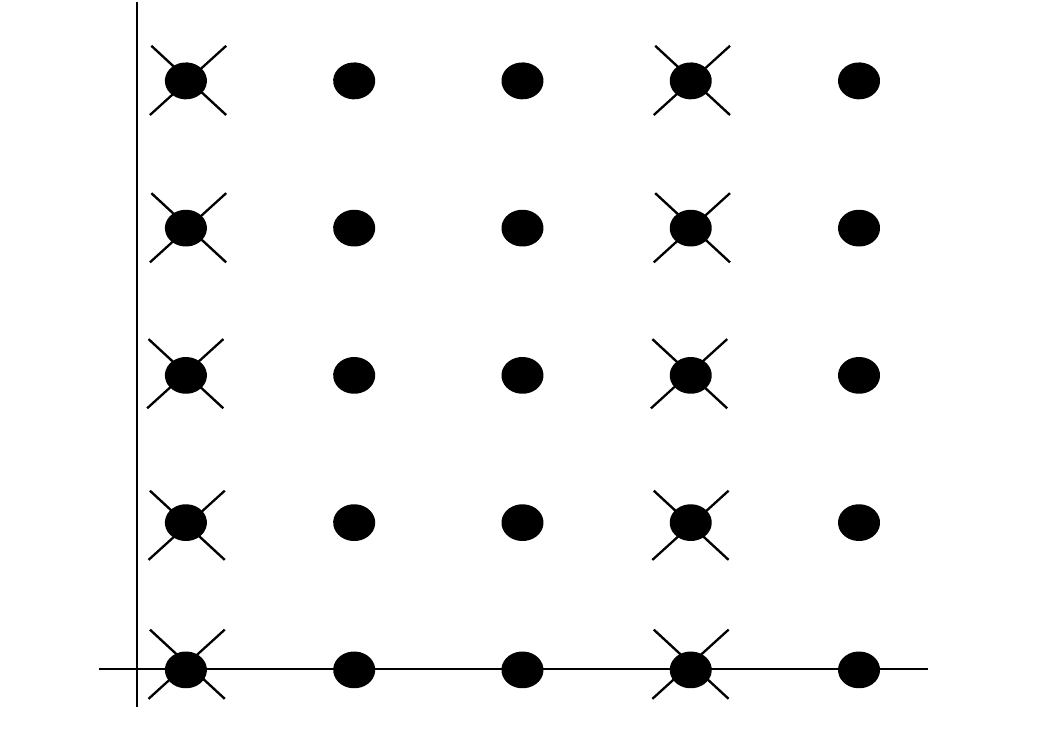}
    \caption{A spatio-temporal sampling set for a dynamical sampling scheme with $D = \ZZ_5$ and $X = \{0,3\}$. Crosses correspond to the sampling locations.} 
  \end{center}
\end{figure}

Figure \ref{Fig1} gives an example of a spatio-temporal sampling set for a finite domain.
More general problems encountered in applications can also be stated. For example, the operators $A_t$ may be unknown or only partially known, and  the sampling set $X$ may be a function of $t$ so that the samples are given by $\{f(X)$, $f_{t_1}(X_1), \dots, f_{t_N}(X_N)\}$. This is a natural problem for wireless sensor networks (WSN), where a large number of sensor nodes are deployed over a physical region to monitor a physical phenomenon such as temperature, pollution concentration, or pressure.  In \cite{ASSC02}, numerous examples of applications of WSN are given in  military, environmental, health, and home and commercial areas. Although several approaches for the reconstruction of signals from WSN samples have emerged recently \cite
{CKL08, RM09Clust, RM09, RM10, RMG12}, most of them do not take into account the evolutionary nature of the sampled processes. A different approach that does exploit the evolutionary aspect of these problems has been proposed and studied in \cite{HRLV10,LV09,RCLV11} and inspired our current research. 

Although dynamical sampling aims to recover a function from samples, it differs from standard sampling problems since it is not only the function $f$ that is sampled but also its various states at different times ($\{t_0, t_1,\dots,t_N\}$). Moreover, it is assumed that at any fixed time $t_i$ there are not enough samples to recover the function $f$  or its state $f_{t_i}$. Although dynamical sampling is an inverse problem, it differs from the typical  inverse problems in which $f_T(X)=(A_Tf)(X)$ is known at a single time $T$. %and we wish to find $f$.    

In this paper we will concentrate on a few special cases of the general dynamical sampling problem.  In particular, we will assume that the initial function $f$ that we want to recover belongs to $\ell^2(\ZZ)$ or  a shift invariant space $V(\phi)$ described below. Furthermore, we will assume that 
the family of operators $A_t$ acting on the initial state  $f$ is spatially  invariant, i.e.,  it is independent of (the absolute) position. This means that for each fixed $t$ we have  $A_tf=a_t\ast f$, that is $A_t$ is a convolution operator. We also assume time invariance in the form $A_{t_1+t_2}=A_{t_1}A_{t_2}$. Additional assumptions on the sampling set $X$ will also be made.  These assumptions allow us to use Fourier techniques and simplify some of the calculations. 

\subsection {Organization}
The paper is organized as follows. Section \ref{Dysamell2} is devoted to necessary and sufficient conditions for solvability of the dynamical sampling problem and has three subsections. The first of these states dynamical sampling results in $\ell^2(\ZZ)$ and the second states results in shift invariant spaces. The last subsection contains the proofs of the results. Section \ref{Stab} deals with the estimates for the reconstruction error in the presence of additive white noise. As in the previous section, we first state the theorems and then provide the proofs in a separate subsection.

\section {Dynamical sampling in $\ell^2(\ZZ)$ and shift invariant spaces.}
\label {Dysamell2}

By $f\in\ell^2(\ZZ)$ we model an unknown spatial signal at time $t=0$. Let $a\in\ell^2(\ZZ)$ represent the kernel of an evolution operator so that the signal at time $t=n$ is given by $a^n\ast x = (\underbrace{a\ast \hdots \ast a}_{n})\ast\,x.$ By $S_m: \ell^2(\ZZ) \rightarrow \ell^2(\ZZ)$ we denote the operator of subsampling by a factor of $m$ so that $(S_mz)(k)=z(mk)$. 
The dynamical sampling problem under these assumptions can  be stated as follows: 

{\it Under what conditions on $a$, $m$, and  $N$ can a function $f \in \ell^2(\ZZ)$ be recovered from the samples $\{S_mf,S_m(a\ast f), \dots, ,S_m(a^{N}\ast f)\}$ of $f$, or, equivalently, from $\{f(X),(a\ast f)(X), \dots, ,(a^{N}\ast f)(X)\}$, $X=m\ZZ$}? 

If we let $y_n = S_m(a^{n-1}\ast f)$, $n = 1, \hdots N$, 
we can rephrase the problem by writing it in the form:
\begeq
\label {OPversion}
\mathbf y =   \mathbf Af,
\eq
where $\mathbf A$ is the operator from  $\ell^2(\ZZ)$ to $\left(\ell^2(\ZZ)\right)^N$, such that $\mathbf y = (y_1,\dots, y_N) =\big(S_mf,S_m(a\ast f), \dots, S_m(a^{N-1}\ast f)\big)$.

In order to stably recover $f$, the operator $\mathbf A$ must have a bounded left inverse. This means that there must exist an operator $\mathbf B$  from $\left(\ell^2(\ZZ)\right)^N$ to $\ell^2(\ZZ)$ such that $\mathbf B \mathbf A=I$. In particular $\mathbf A$ must be injective and its range, $\text {ran} \mathbf A$, must be closed. In the theorem below we provide necessary and sufficient conditions for such an inverse to exist in terms of the Fourier transform $\hat a$ of the filter $a\in \ell^2(\ZZ)$. For $a\in\ell^1(\ZZ)$ the Fourier transform is defined on the torus $\TT \simeq [0,1)$ by
\[\hat a(\xi) = \sum_{n\in\ZZ} a(n)e^{-2\pi i n\xi},\ \xi\in\TT.\]

\bt \label{infinitedim} Assume that $\hat{a} \in L^\infty (\TT)$ and define 
\begeq \label {fdNScond} \mathcal A_m(\xi)= \left(
\begin{array}{cccc} 1 & 1 & \hdots &1\\
\hat{a}(\frac{\xi}{m}) & \hat{a}(\frac{\xi+1}{m}) & \hdots & \hat{a}(\frac{\xi+m-1}{m})\\
\vdots & \vdots& \vdots & \vdots \\
\hat{a}^{(m-1)}(\frac{\xi}{m}) & \hat{a}^{(m-1)}({\frac{\xi+1}{m}}) & \hdots & \hat{a}^{(m-1)}(\frac{\xi+m-1}{m})\end{array}\right),
\eq
 $\xi \in \TT$. Then $\mathbf A$ in \eqref {OPversion} has a bounded left inverse for some $N\ge m-1$  
 if and only if there exists $\alpha > 0$ such that the set $ \{ \xi : |\det \mathcal{A}_m(\xi)| < \alpha \}$ has zero measure. Consequently, $\mathbf A$ in \eqref {OPversion} has a bounded left inverse for some $N\ge m-1$ if and only if $\mathbf A$ has a bounded left inverse for all $N\ge m-1$.  
 \et
 Thus, under the conditions of Theorem \ref {infinitedim} a vector $f\in \ell^2(\ZZ)$ satisfying \eqref {OPversion} can be recovered in a stable way from the measurements $y_n, n = 1, \hdots, N$, for any $N\ge m-1$. We shall  see in the proof that in the case $N=m-1$ the operator $\mathbf A$ is, in fact, invertible and not just left invertible. For the case $N< m-1$, the operator  $\mathbf A$ is not injective and hence no recovery of $f$ is possible. We also note  that if a signal $f$ cannot be recovered from the dynamical samples in Theorem \ref{infinitedim} then taking additional samples at the same spatial locations will not help. The same phenomenon was observed  in \cite{LV09}.
 
 In the special case when $\hat{a}$ is continuous on $\TT$, $|\det \mathcal{A}_m(\xi)|$ is a continuous function over the compact set $\TT$. Therefore,  an $\alpha$ in Theorem \ref{infinitedim} exists if and only if $|\det \mathcal{A}_m(\xi)| \neq 0$ for all $\xi \in \TT$.  We capture this fact in the corollary below.

\bc\label{cor22}
Suppose $\hat{a} \in C(\TT)$. Then  $\mathbf A$ in \eqref {OPversion} has a bounded left inverse for some (and, hence, all)  $N\ge m-1$ 
if and only if $ |\det \mathcal{A}_m(\xi)| \neq 0$  for all $\xi \in \TT$.
\ec

Although Theorem \ref {infinitedim} gives necessary and sufficient conditions on   convolution operators on $\ell^2(\ZZ)$ for this special case of dynamical sampling problem to be solvable, many typical operators encountered in physical systems or in applications do not satisfy these conditions.  
For example, a typical convolution operator  
is such that $\hat{a}$ is real, symmetric, continuous, and strictly decreasing on $[0, \frac 12]$. The following corollary shows that the dynamical sampling problem cannot be solved in this case without additional samples.
\bp \label {ASingularities}
If $\hat{a}$ is real, symmetric, continuous, and strictly decreasing on $[0, \frac 12]$, then $\mathcal A_m(\xi)$ is singular if and only if $\xi \in \left\{0,\frac 12\right\}$.
\ep

\begin{figure}\label{fig2}
  \begin{center}
  \label {SampZAdd}
    \includegraphics{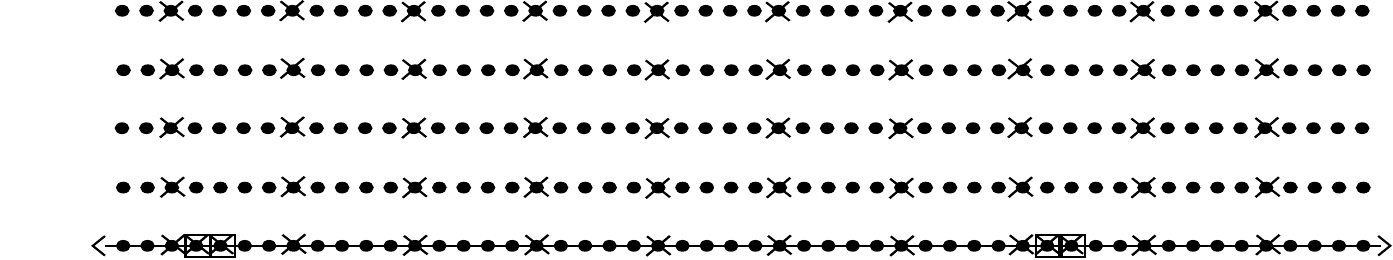}
    \caption{An example of a stable sampling scheme in Theorem \ref{AddSample} with $m = 5$ and $n = 7$.
    The  sampling locations are marked by crosses and  the extra samples at $t=0$ are marked as crosses inside squares.} 
  \end{center}
\end{figure}

Because $\mathcal A_m(0)$ and $\mathcal A_m(\frac 12)$ are not invertible, we cannot solve \eqref {OPversion}. To make reconstruction possible in this case, the sampling set needs to be modified or expanded. This can be done in the following way. Let $T_c$ be the operator that shifts a vector $z\in \ell^2(\ZZ)$ to the right by $c$ units so that $T_cz(k) = z(k-c)$. Let also $S_{mn} T_c$ represent shifting by $c$ and then sampling by $mn$ for some $n\in\NN$.
\bt \label{AddSample}
Suppose $\hat{a}$ is real, symmetric, continuous, and strictly decreasing on $[0, \frac 12]$, $n$ is odd, and $\Omega = \{1, \hdots, \frac{m-1}{2}\}$. Then the extra sampling given by $\left\{S_{mn}T_c\right\}_{c\in\Omega}$ provides enough additional information to stably recover $f$. 
\et
Figure \ref {fig2}  illustrates a sampling set for stable reconstruction.

\brem Note that the extra samples in Theorem \ref{AddSample} needed to recover $f$ are chosen as  $S_{mn} T_cf$. However, it may be natural  to also include the samples on $X_c=mn\ZZ+c$ at $t=1,\dots, N$ for each $c\in \Omega$. In fact, if we have the samples of $f, Af, \dots, A^Nf$ on 
$X\cup\left(\bigcup\limits_{c\in \Omega} X_c\right)$, we can  expect  the recovery process to be more stable. 
\erem

\brem
\label{omega}
Theorems \ref{infinitedim} and \ref{AddSample} parallel the finite dimensional results we obtained in \cite{ADK13}. For example, one can use more complicated choices for $\Omega \subset \{1, \hdots, mn-1\}$ in Theorem \ref{AddSample}, and the admissible choices are determined by the same equivalence relations as in the finite dimensional case (see \cite{ADK13} for more details). Moreover, some of the the methods we use for obtaining stability results in Section \ref{Stab} are the same as in \cite{ADK13}. There are, however, subtle but important differences in the infinite dimensional case. For example, in Theorem \ref{AddSample}, the dynamical samples  without the samples in the extra sampling set $\Omega$ still form a uniqueness set (the operator $\mathbf A$ has a trivial kernel). The latter was not the case in \cite{ADK13}.
 \erem

\subsection {Dynamical Sampling in Shift-Invariant Spaces}

Signals are not always modeled by $\ell^2(\ZZ)$. For example, analog functions are often assumed to belong to a Shift-Invariant Space (SIS) $V(\phi)$, $\phi\in L^2(\RR)$, defined by
\begin {equation}
\label {SIS}
V(\phi) = \left\{  \sum_{k \in \ZZ} c_k \phi(\cdot-k) : (c_k)_{k \in \ZZ} \in \ell^2(\ZZ)\right\}.
\end{equation}
The dynamical sampling problem in shift-invariant spaces is to reconstruct the function $f\in V(\phi)$ from the coarse samples $\{g_0 = S(\Omega_0)f,\ g_n = S_mA^{n-1} f, \ n=1,\dots, N\}$, where  $\Omega_0$ is a ``small'' and possibly empty extra sampling set. Here $S_mg=g(m\cdot)$, $g \in L^2(\RR)$. Although all separable Hilbert spaces are isometrically isomorphic,  the dynamical sampling problem in SIS is not always reducible to that in $\ell^2(\ZZ)$. The reason for this phenomenon is that a convolution operator $A$ acting on a function $f \in V(\phi)$ does not necessarily result in a function $a\ast f$ that belongs to $V(\phi)$. For the case of ($\frac12$)-bandlimited functions $V(\sinc)$, we have that $a\ast f \in V(\sinc)$ for any $ f \in V(\sinc)$, and, in this case, the dynamical sampling in $V(\sinc)$ does reduce to that in $\ell^2(\ZZ)$. The reduction is done in the following way. Let $\hat a \in L^\infty (\RR)$ and define $\hat b =\hat a\chi_{[\frac 1 2, \frac 1 2]}$ and $x=f|_\ZZ$. Then the maps $ f \mapsto  x$ and $ a \ast f \mapsto  b \ast x$ from $V(\sinc)$ to $\ell^2(\ZZ)$ are isometric isomorphisms;  $b \in \ell^2(\ZZ)$ is the inverse discrete Fourier transform of $\hat b \in L^2(\mathbb T)$.  Thus,  solving the dynamical sampling problem in $V(\sinc)$ with a convolution operator defined by a filter $a$ such that $\hat a \in L^\infty (\RR)$ is equivalent to solving the corresponding dynamical sampling problem for $x=f|_\ZZ$ in $\ell^2(\ZZ)$ with the convolution operator defined by the filter $b \in \ell^2(\ZZ)$. 

There are other SIS for which the dynamical sampling problem is reducible to that in $\ell^2(\ZZ)$. 
Necessary and sufficient conditions for the simple reduction similar to the one described above are presented in \cite{AADP13}.
In particular, if  $\phi$ belongs to the Wiener amalgam space $W_0(L^1)=W(C, \ell^1)$ \cite{AG01},  $\{ \phi(\cdot-k): k \in \ZZ\}$ forms a Riesz basis for $V(\phi)$, and  $\sum_k\hat \phi(\xi+k)\ne 0$, then any of the three equivalent conditions below are sufficient for the reduction to the $\ell^2(\ZZ)$ case:
\begin{enumerate}
\item $a\ast \phi \in V(\phi)$;
\item $a\ast V(\phi) \subseteq V(\phi)$;
\item There exists a function $\hat b\in L^2\left[0,1\right]$ such that for every $k\in\ZZ$, 
$$\hat{a}(\xi+k)\hat{\phi}(\xi+k)=\hat b(\xi)\hat{\phi}(\xi+k) \quad a.e. \in [0,1].$$
\end{enumerate}

When the dynamical sampling problem is not reducible to the $\ell^2(\ZZ)$ case, a similar approach can be followed. Let $f = \sum_k c_k \phi(\cdot - k)\in V(\phi)$, $a^j := a*a*...*a$ ($j-1$ convolutions),
$\phi_j := a^j*\phi, \text{ and } \; \Phi_j := \phi_j |_{\ZZ}$.

Letting $(y_l)(k):=\Big(S_m ( a^l * f)\Big)(k)$, $ k \in \ZZ$, $l=0,\dots,m-1$, and using calculations similar to the ones described below for the case of $\ell^2(\ZZ)$, we get
 \[ \left( \begin{array}{c}
 \hat{y}_0 (\xi)   \\
 \hat{y}_1 (\xi) \\
 . \\
 .\\
 .\\
 \hat{y}_{m-1} (\xi)\end{array} \right) = \left( \begin{array}{cccc}
 \hat{\Phi}_0  (\frac{\xi}{m}) & \hat{\Phi}_0  (\frac{\xi+1}{m}) & ... & \hat{\Phi}_0  (\frac{\xi+m-1}{m}) \\
 \hat{\Phi}_1  (\frac{\xi}{m}) & \hat{\Phi}_1  (\frac{\xi+1}{m}) & ... & \hat{\Phi}_1  (\frac{\xi+m-1}{m})\\
. & . & .  & .\\
. & . & .  & .\\
. & . & .  & .\\
 \hat{\Phi}_{m-1}  (\frac{\xi}{m}) & \hat{\Phi}_{m-1}  (\frac{\xi+1}{m}) & ...  & \hat{\Phi}_{m-1}  (\frac{\xi +m -1 }{m}) \end{array} \right)\left( \begin{array}{c}
  \hat{c}  (\frac{\xi}{m})   \\
  \hat{c} (\frac{\xi +1}{m}) \\
  . \\
  .\\
  .\\
  \hat{c} (\frac{\xi+m-1}{m})\end{array} \right).\]
In short notation, we have
\begin{equation}\label{eqyAc}
\hat{\textbf{y}}(\xi) = \widetilde{\cA}_m (\xi)  \hat {\textbf{c}}_m (\xi).
\end{equation} 
It is now easy to see how to get the results corresponding to Theorem \ref  {infinitedim}  and Corollary \ref {cor22} for the case of SIS. In particular, if $\phi \in W_0(L^1)$ and $a \in W(L^1)$, then $\widehat{\Phi}_j \in C (\TT)$ for $j=1,\dots,m$, and a vector $f\in V(\phi)$ can be recovered in a stable way 
 from the measurements $y_n$,  $n =0,  \hdots, m-1,$ if and only if  $\det \widetilde{\mathcal{A}}_m(\xi)\ne 0$  for all $\xi\in [0,1]$. 
 We refer to \cite {AADP13} for more details on the subject.

As in the $\ell^2(\ZZ)$ case, there are many situations in practice for which the hypotheses of Theorem \ref{infinitedim} are not satisfied and additional samples are needed. For example,
when both $\hat{a}$ and $\hat{\phi}$ are real and symmetric,  the functions $\hat{\Phi}_j$ are also real and symmetric,  forcing $\widetilde{\cA}_m(\xi)$ to be singular at $\xi=0,\frac 12$,  as well as possibly other values of $\xi$. In 
special cases, the number of additional samples and their locations 
may be determined from
the Theorem \ref {addsamp2} below. 

As before, $T_c$, $c\in\ZZ$, are the operators that shift a vector in $\ell^2(\ZZ)$ to the right by $c$ units so that $T_cz(k) = z(k-c)$, and $S_{mn} T_c$ represent shifting by $c$ followed by sampling on $mn\ZZ$ for some positive integer $n$. 

\begin{thm}\label{addsamp2}
Suppose $\widetilde{\cA}_m(\xi)$ is singular only when $\xi \in \{\xi_i\}_{i\in I}$ with $|I| < \infty$.  Let $n$ be a positive integer such that $| \xi_i - \xi_j | \neq \frac{k}{n}$ for any $i,j \in I$ and $k \in \{1, \hdots, n-1\}.$  Then the extra samples given by $\left\{S_{mn}T_c\right\}_{c\in\{1, \hdots, m-1\}}$ provide enough additional information to stably recover any $f \in V(\phi)$. \end{thm}

\brem The finite nature of $I$ guarantees the existence of an $n$ satisfying the conditions of Theorem \ref{addsamp2}. The proof of Theorem \ref {addsamp2} is similar but simpler than that of Theorem \ref {AddSample} and  will be omitted (see also \cite{AADP13}). 
\erem

\subsection {Proofs for Section \ref {Dysamell2}} ${}$\\
The following Lemma is useful for proving Theorem \ref {infinitedim}. 
\bl \label{OpNorm}
Suppose $\mathscr A: (L^2(\TT))^m \to (L^2(\TT))^n$ is defined by $(\mathscr Ax)(\xi) = A(\xi)x(\xi)$ where the map $\xi \mapsto A(\xi)$ from $\TT$ to the space of $n\times m$ matrices $\mathcal M^{n m}$ is measurable. Then $\|\mathscr A\|_{op} = \ssup_{\TT} \| A(\xi)\|_{op}$. 
\el
\bpf %[Proof of Lemma \ref{OpNorm}.]
The proof is standard. 

Suppose $\ssup_{\TT} \|A(\xi)\|_{op} = \alpha < \infty $ and let $z \in (L^2(\TT))^m$ be such that $\|z\|_{(L^2(\TT))^m} = 1$. Then  it is easy to see that $\|\mathscr A z\|_{(L^2(\TT))^m}^2 \le \alpha^2$
and,
therefore, $\|\mathscr A\|_{op} = \sup\limits_{\|z\|_{(L^2(\TT))^m} = 1} \|\mathscr Az\|_{(L^2(\TT))^n} \leq \alpha.$

To prove the opposite inequality, let $\epsilon > 0$ and  $B= \{\xi:\|A(\xi)\|_{op} \ge \alpha - \epsilon\}$. Using  the singular value decomposition we write $A$ as $A(\xi) = U(\xi)\Sigma(\xi)V^\ast(\xi)$
where $U(\xi)$ is an $n\times n$ unitary matrix, $\Sigma(\xi)$ is an $n\times m$ matrix with nonnegative, real entries on the diagonal, and $V(\xi)$ is an $m\times m$ unitary matrix. We assume that the diagonal entries, $s_i(\xi)$ of $\Sigma(\xi)$, called singular values of $A(\xi)$, are listed in descending order. Then $\|A(\xi)\|_{op} =\sqrt{ s_1(\xi)}.$ Let $v_1(\xi)$ be the first column vector of $V(\xi)$, and define
\begeq
z(\xi) = \frac {1}{\sqrt{|B|}}\chi_B(\xi) v_1(\xi),
\eq
where $\chi_B$ is the characteristic function of the set $B$. Since the function $z$ is measurable  
we get
\begeq
\|\mathscr A z\|^2 = \frac {1}{|B|} \int_B |A(\xi)z(\xi)|^2 d\xi = \frac {1}{|B|} \int_B |\sigma_1(\xi)|^2 d\xi \ge (\alpha - \epsilon)^2.
\eq
Thus, $\|\mathscr A\|_{op} \geq \alpha.$

Assume now that $\ssup_{\TT} \|A(\xi)\|_{op} = \infty.$ Fix $N > 0$. Then the set $B =  \{\xi:\|A(\xi)\|_{op} \ge N\}$ has positive measure. Repeating the process above, we find a function $z_N$ of unit norm in $(L^2(\TT))^m$ such that $\|\mathscr Az_N\|_{(L^2(\TT))^n} \ge N$. Since $N$ was arbitrary, we conclude that $\| \mathscr A \| _{op}= \infty.$ In particular, $\mathscr A$ is a bounded operator if and only if $\ssup_{\TT} \| A(\xi)\|_{op} < \infty.$
\epf

Assume now that the matrix $A(\xi)$ in the theorem above has a bounded left inverse for almost every $\xi$, denoted $A^\ell(\xi)$. Then a left inverse $\mathscr A^{\ell}$ can be defined on the range of $\mathscr A$ by $(\mathscr A^{\ell}y)(\xi) = A^\ell(\xi)y(\xi).$ However, in this case, $\mathscr A^{\ell}$ will be a bounded operator if and only if the range of of $\mathscr A$ is closed.

\bpf[Proof of Theorem \ref{infinitedim}.]

Using the fact that 
$$\sum\limits_{l=0}^{m-1}e^{\frac{i2\pi l}{m}j} = \begin{cases} m, & j=0 \text{ mod } m \\ 0, & \text{otherwise} \end{cases},$$ we get the Poisson summation formula
\begeq \label{DSformula} (S_mz)^\wedge(\xi) =  \frac{1}{m}\sum\limits_{l=0}^{m-1} \hat{z}(\frac{\xi+l}{m}), \quad \xi \in \TT, \ z \in \ell^2(\ZZ).\eq 
Let $G :  L^2(\TT) \rightarrow (L^2(\TT))^m$ be given by 
\begeq \label{gismo}
(Gz)(\xi) = \frac {1}{\sqrt{m}} \left(z(\frac{\xi}{m}), z(\frac{\xi+1}{m}), \hdots, z(\frac{\xi+m-1}{m})\right)^T.
\eq

Taking the Fourier transform of \eqref {OPversion} we get
\begeq \label {Fouriertradeoff}
 m\left( \begin{array}{c} \hat y_1(\xi) \\ 
\hat y_2(\xi)\\ \vdots \\
 \hat y_N(\xi) \end{array} \right) = \mathcal A_m(\xi)
\left( \begin{array}{c} \hat{f}(\frac{\xi}{m})\\
\hat{f}(\frac{\xi+1}{m})\\
\vdots\\
\hat{f}(\frac{\xi+m-1}{m})\end{array}\right), 
\eq
or, using a more compact notation, 
\begeq \label {Fouriertradeoff1}
\vecy (\xi) = \frac1m \mathcal A_m(\xi) \vecx (\xi) ,
\eq
where $\vecx =\sqrt m\, G\hat f$. Define the operator $\mathscr A : (L^2(\TT))^N \rightarrow (L^2(\TT))^m$ by $(\mathscr A \vecx)(\xi) = \mathcal A_m(\xi) \vecx (\xi)$, where

\begeq \label {AMN} \mathcal A_m(\xi)= \left(
\begin{array}{cccc} 1 & 1 & \hdots &1\\
\hat{a}(\frac{\xi}{m}) & \hat{a}(\frac{\xi+1}{m}) & \hdots & \hat{a}(\frac{\xi+m-1}{m})\\
\vdots & \vdots& \vdots & \vdots \\
\hat{a}^{(N-1)}(\frac{\xi}{m}) & \hat{a}^{(N-1)}({\frac{\xi+1}{m}}) & \hdots & \hat{a}^{(N-1)}(\frac{\xi+m-1}{m})\end{array}\right).
\eq
Since $G$ is an isometric isomorphism, the signal $f$ can be recovered from $\vecy$ in a stable way if and only if the operator $\mathscr A$ has a bounded left  inverse. 
Now it is easy to see that the operator $\mathscr A$ has a bounded left inverse for some $N\ge m-1$, if and only it has a bounded (left) inverse for $N= m-1$. The latter happens  if and only if there exists $\alpha > 0$ such that the set $ \{ \xi : |\det \mathcal{A}_m(\xi)| < \alpha \}$ has zero measure. 
\epf

\bpf [Proof of Proposition \ref {ASingularities}]
The Vandermonde matrix $\mathcal A_m(\xi)$ in \eqref{fdNScond} is singular if and only if two of its columns coincide. Suppose the $j$-th and $l$-th columns coincide and  $j < l$. Then $\hat{a}(\frac {\xi +j}{m}) = \hat{a}(\frac {\xi +l}{m})$. The symmetry and monotinicity conditions on $\hat{a}$ imply that $\frac {\xi +j}{m} = 1- \frac {\xi +l}{m}$.  Then $\xi = \frac{m-j-l}{2}$. Observing that $m-j-l \in \ZZ$ and $\xi \in \TT$, we conclude that $\xi \in \left\{0,\frac 12\right\}$. 
\epf

\bpf [Proof of Theorem \ref {AddSample}]
We begin with a few useful formulas and notation. Combining the identity \eqref{DSformula} with the identity
\begeq
(T_c f)^\wedge(\xi) = e^{-i2\pi c \xi} \hat{f}(\xi)
\eq
we get
\begin{eqnarray*}
(S_{mn} T_c f)^\wedge (n\xi) &=& \frac{1}{mn} e^{\frac{-i2\pi c\xi}{m}} \sum\limits_{l=0}^{mn-1} e^{\frac{-i2\pi c l}{mn}} \hat{f}(\frac{\xi}{m}+\frac{l}{mn})\\
&=&\frac{1}{mn} e^{\frac{-i2\pi c\xi}{m}} \sum\limits_{k=0}^{n-1} e^{\frac{-i2\pi c k}{mn}}\sum\limits_{j=0}^{m-1} e^{\frac{-i2\pi cj}{m}}\hat{f}(\frac{\xi+j}{m}+\frac{k}{mn}).
\end{eqnarray*}
Using the notation of \eqref{Fouriertradeoff1} and defining the row vector
\begeq
\vecu_c(k) = e^{\frac{-i2\pi c k}{mn}}(1, e^{\frac{-i2\pi c}{m}}, e^{\frac{-i4\pi c}{m}}, \hdots, e^{\frac{-i2\pi c(m-1)}{m}}),
\eq
we have
\begeq
(S_{mn}T_c f)^\wedge (n\xi) = \frac{1}{mn} e^{\frac{-i2\pi c\xi}{m}} \sum\limits_{k=0}^{n-1} \vecu_c(k) \vecx(\xi + \frac{k}{n}),
\eq
where $\vecx(\xi )=\left(\hat f(\frac\xi m), \hat f(\frac{\xi+1} m), \dots, \hat f(\frac{\xi+m-1} m)\right)^T = \sqrt m\, G\hat f$ as before.

We  consider an initial extra sampling set $\Omega = \{1, \hdots, \frac {m-1} 2\}$. Combining the dynamical  samples with the additional initial samples we have 
\begeq 
\label {addini}
m\left( \begin{array}{c} ne^{\frac{i2\pi \xi}{m}}(S_{mn}T_1 x)^\wedge (n\xi)\\
\vdots\\
ne^{\frac{i2\pi \xi(m-1)}{2m}}(S_{mn}T_{\frac {(m-1)} 2} x)^\wedge (n\xi)\\
\vecy(\xi)\\
\vecy(\xi+\frac1n)\\
\vdots\\
\vecy(\xi+\frac{n-1}{n}) \end{array} \right) = \A(\xi)
\left(\begin{array}{c} \vecx(\xi)\\
\vecx(\xi + \frac1n)\\
\vdots\\
\vecx(\xi + \frac{n-1}{n})\end{array}\right),
\eq
where $\A$ is given by
\begeq
\label {Aextended}
\A(\xi) = \left( \begin{array}{cccc}
\vecu_1(0) & \vecu_1(1) & \hdots & \vecu_1(n-1)\\
\vdots & \vdots & \ddots & \vdots\\
\vecu_{\frac {m-1} 2}(0) & \vecu_{\frac {m-1} 2}(1) & \hdots & \vecu_{\frac {m-1} 2}(n-1)\\
\mathcal{A}_m(\xi) & 0 & \hdots & 0\\
0 & \mathcal{A}_m(\xi + \frac1n) & \hdots & 0\\
\vdots & \vdots & \ddots & \vdots\\
0 & 0 & \hdots & \mathcal{A}_m(\xi + \frac{n-1}{n})\end{array}\right).
\eq

If $\A(\xi)$ has full column rank, then it has a left inverse. By Lemma \ref{OpNorm} and the fact that $\hat{a}$ is continuous, it suffices to show that the matrix $A(\xi)$ has full rank for every $\xi \in [0, \frac 1n]$; it is not difficult to see that  solving \eqref {addini} for $\xi \in [0,\frac 1 n]$ is sufficient for the recovery of  $\vecx(\xi)$ for all $\xi \in [0,1]$.

First, if $\xi + \frac kn \notin \left\{0, \frac 12 \right\}, k = 0, \hdots n-1,$ the solvability is implied by Proposition \ref{ASingularities}. Next, notice that for a fixed $\xi \in [0,\frac 1 n]$, we have $\xi + \frac kn \in \left\{0, \frac 12 \right\}$ for at most one $k = 0, \hdots n-1.$ This follows from the parity of $n$ ($n$ is assumed to be odd). Therefore, \ for any $\xi\in [0, \frac 1n]$ there is at most one singular block $\mathcal A_m$ in $\A(\xi)$. This allows us to consider the singularities of $\mathcal A_m(0)$ and $\mathcal A_m(\frac 12)$ separately. 

Because of the block diagonal structure of the lower portion of $\A(\xi)$, we can focus only on showing that the additional samples eliminate any singularities created by $\mathcal A_m(0)$ and $\mathcal A_m(\frac 12)$.

For a fixed $k = 0, \hdots, \frac {m-1} 2$, we define the $\frac {(m-1)} 2 \times m$ matrix
\begeq
U_k = \left(\begin{array}{c} \vecu_1(k)\\
\vdots\\
\vecu_{\frac {m-1} 2}(k)\end{array}\right).
\eq
Since
\begin{eqnarray*}
\left\langle U_k (c,.), U_k (d,.) \right\rangle &=& \sum_{j=0}^{m-1} e^{\frac{-i2\pi c k}{mn}}e^{\frac{-i2\pi cj}{m}}e^{\frac{i2\pi d k}{mn}}e^{\frac{i2\pi dj}{m}}\\
&=& e^{\frac{-i2\pi (c-d) k}{mn}}\sum_{j=0}^{m-1} e^{\frac{-i2\pi j (c-d)}{m}}\\
&=& \begin{cases} me^{\frac{-i2\pi (c-d) k}{mn}}, & (c-d) = 0 \:mod\, m\\ 0, & otherwise \end{cases},
\end{eqnarray*}
%Thus, 
the rows of the matrix $U_k$ form an orthogonal set, and we conclude that it has full rank. 
Next, we show that $\left(\begin{array}{c} U_k\\
\mathcal A_m(\xi + \frac kn) \end{array}\right)$ has a trivial kernel and, hence, full rank. 
A vector is in $\ker\left(\begin{array}{c} U_k\\
\mathcal A_m(\xi + \frac kn) \end{array}\right)$ if and only if it is in the kernels of both $\mathcal A_m(\xi + \frac kn)$ and $U_k$. Therefore, we  only need to consider $\xi+\frac k n\in \{0,\frac 12\}$.

Under the  conditions of Proposition \ref {ASingularities}, we can completely characterize the kernels of $\mathcal A_m(0)$ and $\mathcal A_m(\frac 12)$. For simplicity, we assume $m$ is odd and begin indexing the columns of $\mathcal A_m(\xi)$ at zero. When $\xi = 0$, the $l$-th column of the Vandermonde matrix  $\mathcal A_m(0)$ is found by evaluating $\hat{a}$ at $\frac {l}{m}$. By the symmetry and $1$-periodicity of $\hat{a}$, we have $\hat{a}(\frac {j}{m}) = \hat{a}(\frac {m-j}{m})$. Therefore, the $j$-th and $(m-j)$-th columns of $\mathcal A_m(0)$ coincide, and the kernel of $ \mathcal A_m(0)$ has dimension $\frac {m-1} 2$. 
Similarly, the $j$-th and $(m-j-1)$-th columns of $\mathcal A_m(\frac 12)$ coincide for $j = 0, \hdots, \frac{m-3}{2}$ and and the kernel of $ \mathcal A_m(\frac 1 2)$ also has dimension $\frac {m-1} 2$.

The vector $\vecv_j$ with a $1$ in the $j$-th position, a $(-1)$ in the $(m-j)$-th position, and zeros elsewhere is in the kernel of $\mathcal A_m(0)$. Since there are exactly $\frac {m-1} 2$ of such vectors, the kernel is their span:
\begeq \label{v_j} \ker \mathcal A_m(0) = \text{span}\left\{\left( \begin{array}{c}0\\
1\\
0\\
\vdots\\
0\\
-1\end{array} \right),
\left( \begin{array}{c}0\\
0\\
1\\
\vdots\\
-1\\
0\end{array} \right),
\hdots,
\left( \begin{array}{c}0\\
\vdots\\
1\\
-1\\
\vdots\\
0\end{array} \right) \right\} = \text{span} \{\vecv_j\}_{j=1}^ {\frac{m-1}{2}}. \eq

Similarly, the $j$-th and $(m-j-1)$-th columns of $\mathcal A_m(\frac 12)$ coincide for $j = 0, \hdots, \frac{m-3}{2}$, and the vector $\vecw_j$ with a $1$ in the $j$-th position, a $(-1)$ in the $(m-j-1)$-th position, and zeros elsewhere is in the kernel of $\mathcal A_m(\frac 12)$. Therefore, 
\begeq \ker \mathcal A_m(\frac 12) = \text{span}\left\{\left( \begin{array}{c}1\\
0\\
\vdots\\
0\\
\vdots\\
0\\
-1\end{array} \right),
\left( \begin{array}{c}0\\
1\\
\vdots\\
0\\
\vdots\\
-1\\
0\end{array} \right),
\hdots,
\left( \begin{array}{c}0\\
\vdots\\
1\\
0\\
-1\\
\vdots\\
0\end{array} \right) \right\} = \text{span} \{\vecw_j\}_{j=0}^ {\frac{m-3}{2}}. \eq
Suppose $\vecx \in \ker \mathcal A_m(0)$.  
Then $\vecx = \sum_{j=1}^{\frac{m-1}{2}} \alpha_j\vecv_j$, where $\vecv_j$ is defined in \eqref{v_j}. We want to know if the equation $0 =  U_k\vecx = \sum_{j=1}^{\frac{m-1}{2}} \alpha_j U_k\vecv_j$ has a unique (trivial) solution. This happens if and only if the matrix $B =  U_k \left( \vecv_1 \hdots \vecv_{\frac{m-1}{2}}\right)$ has full rank.   
Computing the $(c,j)$ entry of $B$, we have 
\begeq
\begin{array}{ccc} B(c,j) &=& e^{\frac{-i2\pi c k}{mn}}(e^{\frac{-i2\pi}{m}cj}-e^{\frac{-i2\pi}{m}c(m-j)})\\
&=& e^{\frac{-i2\pi c k}{mn}}(e^{\frac{-i2\pi}{m}cj}-e^{\frac{i2\pi}{m}cj})\\
&=& 2e^{\frac{-i2\pi c k}{mn}}\sin(\frac{2\pi}{m}cj).
\end{array}
\eq
Note that $\{1, \cos( \frac{2\pi}{m}cj),  \sin(\frac{2\pi}{m}cj): c=1,\dots,\frac{m-1} 2\}$ is the Fourier basis for  $\CC^m$. Thus, $ \{\sin(\frac{2\pi}{m}cj): c=0,\dots,\frac{m-1} 2\}$ are linearly independent in $\CC^m$. Using the fact that $ \{\sin(\frac{2\pi}{m}cj): c=0,\dots,\frac{m-1} 2\}$ are odd functions, it follows that $ \{\sin(\frac{2\pi}{m}cj): c=0,\dots,\frac{m-1} 2\}$ form a basis of $\CC^{\frac {m-1} 2}$. 
Therefore, the $\frac {m-1} 2 \times \frac{m-1}{2}$ matrix $B$ does, indeed, have full rank.

Similarly, for the case $\mathcal A_m(\frac 1 2)$, we consider the matrix $D = U_k \left( \vecw_0 \hdots \vecw_{\frac{m-3}{2}}\right)$. Its entries are
\begeq
D(c,j) = 2e^{\frac{-i2\pi c k}{mn}}e^{\frac{-i\pi}{m}c}\sin(\frac{2\pi}{m}c(2j+1)),
\eq
and, therefore, $D$ has full rank.

Thus,  
the matrix $\A$ has a bounded left inverse for every $\xi\in\TT$ and the theorem is proved.
\epf

\section{stability in the presence  of additive noise}\label{Stab}

In this section, we assume that $\hat{a}$ and $\Omega$ satisfy the hypotheses of Theorem \ref{AddSample} and consider the recovery of the signal $f$ in the presence of additive noise. The minimal extra sampling set $\Omega$ in Theorem \ref{AddSample} allows us to stably recover any signal $f \in \ell^2(\ZZ)$.  In the presence of additive Gaussian white noise, however, any  linear recovery method does not generally reproduce the original function $f$. Under the above hypotheses, the expected discrepancy, $\tilde f -f$, between the recovered function $\tilde f $ and the original function $f$  is controlled by the norm of the operator $\mathbf A_\Omega^\dagger : (L^2(\TT))^{mn+|\Omega|} \rightarrow (L^2(\TT))^{mn}$ defined by $(\mathbf A_\Omega^\dagger y)(\xi) = A_\Omega^\dagger (\xi) y(\xi)$, where $A_\Omega^\dagger(\xi)$ is the Moore-Penrose pseudoinverse of the matrix $A_\Omega(\xi)$ in  
\eqref {ExtraSamples} below. An upper bound for  $\|\mathbf A_\Omega^\dagger\|$  is given in the following theorem.

\bt
\label {stabup}
If $\Omega=\{0, \dots, m-1\}$ and $\hat{a}$ and $n$ satisfy the hypotheses of Theorem \ref{AddSample} then 
\begeq \nonumber
\|\mathbf A_\Omega^\dagger\| \leq m\beta_1(1+ m\sqrt{n-1})
\eq
where $\beta_1 = \max \{n, \underset{\xi \in J}{\ssup} \|\mathcal A_m^{-1}(\xi)\|\}< \infty$,
 $J =  [\frac{1}{4n}, \frac 12 - \frac{1}{4n}] \cup [\frac 12 + \frac{1}{4n}, 1 - \frac{1}{4n}],$ and
 $\mathcal A_m(\xi)$ is defined by \eqref{fdNScond}.
\et

In the following corollaries we give more explicit bounds for the value of $\beta_1$. There, without loss of generality, we assume that $\sup|\hat{a}(\xi)|\leq 1$.

\bc
\label {cor1stabup}
If $\Omega=\{0, \dots, m-1\}$, $\hat{a}$ and $n$ satisfy the hypotheses of Theorem \ref{AddSample}, and $\sup|\hat{a}(\xi)|\leq 1$ then 
\begeq \nonumber
\|\mathbf A_\Omega^\dagger\| \leq m\beta_2(1+ m\sqrt{n-1})
\eq
where $\beta_2 = \max \left\{n, \left(\frac{2}{\delta}\right)^{m-1}\right\}< \infty$,
$\delta = \underset{\begin{smallmatrix} \xi \in J\\j=0, \dots, m-1 \\ j\neq i \end{smallmatrix}}{\min}|\hat a(\frac{\xi+j}{m})-\hat a(\frac{\xi+i}{m})|,$
and $J =  [\frac{1}{4n}, \frac 12 - \frac{1}{4n}] \cup [\frac 12 + \frac{1}{4n}, 1 - \frac{1}{4n}].$
\ec

\bc
\label {cor2stabup}
If $\Omega=\{0, \dots, m-1\}$, $\hat{a}$ and $n$ satisfy the hypotheses of Theorem \ref{AddSample}, $\sup|\hat{a}(\xi)|\leq 1$, and, in addition, $\hat{a} \in C^1(0, \frac 12)$ and the derivative $\hat{a}'$ of $\hat{a}$ is nonzero (and, hence, negative) on $(0, \frac 12)$, then 
\begeq \nonumber
\|\mathbf A_\Omega^\dagger\| \leq m \beta_3(1+ m\sqrt{n-1}),
\eq
where $\beta_3 = \max\left\{n, \left(\frac {4mn}{\gamma}\right)^{m-1}\right\}$, $\gamma = \min\limits_M|\hat{a}'(\xi)|$, and
$M = \left[\frac{1}{4mn}, \frac 12 - \frac{1}{4mn}\right]$. 
\ec

For a Gaussian i.i.d.~additive noise $\mathcal N(0,\sigma^2)$ a reconstruction of $f$ using $\mathbf A_\Omega^\dagger$ will result in an error estimated by $\|f-\tilde f\|\le \|\mathbf A_\Omega^\dagger\|\sigma m^{-\frac12}$. 
The theorem above provides an upper bound for the operator norm $\|\mathbf A_\Omega^\dagger\|$. However, although the upper bound grows to infinity as $n$ or $m$ increases, it is not yet clear that $\|\mathbf A_\Omega^\dagger\|$ deteriorates in this case. The following two results show that, indeed, as $m$ or $n$ increases $\|\mathbf A_\Omega^\dagger\|$ is unbounded and the stability of reconstruction does in fact worsen. 

\bt \label{stablow}
Suppose $\hat{a}$, $n$, and $\Omega$ satisfy the hypotheses of Theorem \ref{AddSample} with $|\Omega| = \frac{m-1}{2}$. Then $\|\mathbf A_\Omega^\dagger\| \geq m \|\mathcal A_m^{-1}(\frac 1n)\|$. 
\et

\bc \label{cor1stablow}
Suppose $\hat{a}$, $n$, and $\Omega$ satisfy the hypotheses of Theorem \ref{AddSample} with $|\Omega| = \frac{m-1}{2}$. Then $\|\mathbf A_\Omega^\dagger\| \rightarrow \infty$ as $n \rightarrow \infty$.
\ec

\begin {rem}
The proof of the theorem shows that
if $\Omega$ is some larger set, that is $|\Omega| > \frac{m-1}{2}$, then  the growth of $\|A^\dagger\|$
may be alleviated. It should also be noted that in practice sampling on $\Omega$ will also likely to be  performed at all times $n=0,\dots,m-1$, rather than just when $n=0$. This may also have the effect of decreasing $\|A^\dagger\|$.  

\end{rem}

\subsection {Proofs of Theorems}
In the beginning, we provide two well-known lemmas that we use in the proofs.

\bl \label{LeftInverseNorm}
Let $A$ be an $m\times n$ matrix with $m>n$ so that the Moore-Penrose left inverse is given by $A^\dagger = (A^*A)^{-1}A^*$. If $A^{\ell}$ is any other left inverse of $A$, then $\|A^\dagger\| \leq \|A^{\ell}\|$. 
\el

\bl \label{PermutationOfRowsAndColumns}
Suppose $A$ is an $m \times n$ matrix with $m>n$, the maps $\sigma: \{1, \hdots m\} \rightarrow \{1, \hdots m\}$ and $\eta: \{1, \hdots n\} \rightarrow \{1, \hdots n\}$ are permutations, and $B$ is an $n \times m$ matrix such that $BA = I_n$. If the matrices $\tilde{A}$ and $\tilde{B}$ are given by $\tilde{A}(i,j):= A(\sigma(i), \eta(j))$ and $\tilde{B}(j,i):= B(\eta(j), \sigma(i))$, then $\tilde{B}\tilde{A} = I_n$ and $\|B\|_{op} = \|\tilde{B}\|_{op}.$
\el

\subsubsection {Proof of Theorem \ref {stabup}}

Similar to the matrix \eqref {Aextended}, the matrix obtained for the additional sampling on $ \Omega=\{1,\dots,m-1\}$ is given by 
\begeq \label{ExtraSamples}
 A_\Omega (\xi) = \left( \begin{array}{cccc}
\frac {1}{mn} \vecu_1(0) & \frac {1}{mn} \vecu_1(1) & \hdots & \frac {1}{mn} \vecu_1(n-1)\\
\vdots & \vdots & \ddots & \vdots\\
\frac {1}{mn} \vecu_{m-1}(0) & \frac {1}{mn} \vecu_{m-1}(1) & \hdots & \frac {1}{mn} \vecu_{m-1}(n-1)\\
\frac 1m\mathcal{A}_m(\xi) & 0 & \hdots & 0\\
0 & \frac 1m\mathcal{A}_m(\xi + \frac1n) & \hdots & 0\\
\vdots & \vdots & \ddots & \vdots\\
0 & 0 & \hdots & \frac 1m\mathcal{A}_m(\xi + \frac{n-1}{n})\end{array}\right).
\eq

In light of Lemma \ref{OpNorm}, a uniform upper bound for $\|A_\Omega^\dagger(\xi)\|$, that is an upper bound independent of $\xi$, provides an upper bound for $\|\mathbf A_\Omega^\dagger\|.$ We choose $\Omega = \{0,1, \hdots, m-1\}$ and $n, m$ to be odd.

We will rearrange the rows and columns of the matrix $A_\Omega(\xi)$ to create a matrix $\tilde{A}_\Omega(\xi)$ for which we can explicitly give a left inverse. By Lemmas \ref{PermutationOfRowsAndColumns} and \ref{LeftInverseNorm}, it suffices to find an upper bound for any left inverse of $\tilde{A}_\Omega(\xi)$.

For a fixed $\xi \in [0, \frac 1n],$ let $k_0$ be such that $\xi+\frac{k_0}{n}$ is the closest point of $\{\xi+\frac kn\}_{k = 0, \dots n-1}$ on the torus to a singularity of $\mathcal A_m$. Specifically, if $\xi \in [0, \frac {1}{4n}),$ then $k_0 = 0$; if $\xi \in [\frac {1}{4n}, \frac {3}{4n}),$ then $k_0 = \frac{n-1}{2}$; and if $\xi \in [\frac {3}{4n}, \frac {1}{n}],$ then $k_0 = n-1$.   We see that 
\begeq \label{MinDist}
\min_{\begin{smallmatrix}k=0,\hdots, n-1\\ k \neq k_0\end{smallmatrix}} \left\{\dist(\xi+\frac kn,0), \dist(\xi+\frac kn,\frac 12), \dist(\xi+\frac kn,1) \right\}\geq \frac{1}{4n}.
\eq
In other words, for $k \neq k_0,$ and $\xi \in [0, \frac 1n],$ we have $\xi + \frac kn \in J$ where $J=J(n)$ is defined by
\begeq
\label {Jdef}
J =  [\frac{1}{4n}, \frac 12 - \frac{1}{4n}] \cup [\frac 12 + \frac{1}{4n}, 1 - \frac{1}{4n}].
\eq

By rearranging the columns and rows of the matrix $A_\Omega$ so that it has the form $\tilde A_\Omega$ below, we are able to explicitly define a left inverse that is independent of $\mathcal A_m(\xi + \frac{k_0}{n})$. We write
\begeq \label {GenMat} \tilde  A_\Omega (\xi) = \left( \begin{array}{cccc}
\frac {1}{mn} \vecu_0(k_0) & \frac {1}{mn} \vecu_0(k_1) & \hdots & \frac {1}{mn} \vecu_0(k_{n-1})\\
\vdots & \vdots & \ddots & \vdots\\
\frac {1}{mn} \vecu_{m-1}(k_0) & \frac {1}{mn} \vecu_{m-1}(k_1) & \hdots & \frac {1}{mn} \vecu_{m-1}(k_{n-1})\\
\frac 1m\mathcal{A}_m(\xi + \frac{k_0}{n}) & 0 & \hdots & 0\\
0 & \frac 1m\mathcal{A}_m(\xi + \frac{k_1}{n}) & \hdots & 0\\
\vdots & \vdots & \ddots & \vdots\\
0 & 0 & \hdots & \frac 1m\mathcal{A}_m(\xi + \frac{k_{n-1}}{n})\end{array}\right).
\eq
 in the block form 
\begeq \label{GenMatBlock}
\tilde A_\Omega(\xi) =\frac 1m \left(\begin{array} {cc} mU_{k_0} & mQ\\ \mathcal A_m(\xi+\frac{k_0}{n}) & 0 \\ 0& D(\xi)
\end {array}\right),
\eq
where $U_{k_0} = \left( \begin{array}{c} \frac {1}{mn} \vecu_0(k_0)\\ \frac {1}{mn} \vecu_1(k_0)\\ \vdots \\\frac {1}{mn} \vecu_{m-1}(k_0) \end{array}\right)$, and $D(\xi)$ is a $m(n-1)\times m(n-1)$ block diagonal matrix with $\mathcal A_m(\xi + \frac kn), k = 0, \hdots, n-1, k \neq k_0$, on the main diagonal. Then, a left inverse is given by
\begeq \label {GenMatINv}A_\Omega^\ell(\xi) =m\left(\begin{array} {ccc} \frac 1m U_{k_0}^{-1} & 0& - U_{k_0}^{-1} QD^{-1}(\xi)\\0 &0 & D^{-1}(\xi)
\end {array}\right),
\eq
and we easily compute that
\begeq
\label {est1}
\|A_\Omega^\ell(\xi) \| \le m(\max\{ \|\frac 1m U_{k_0}^{-1}\|, \|D^{-1}(\xi)\|\}+  \|U_{k_0}^{-1}\|\|D^{-1}(\xi)\|\|Q\|).
\eq
Since $D$ is a block diagonal matrix, we have
\begeq
\|D^{-1}(\xi)\|=\max_{k\neq k_0} \left\{ \|\mathcal A_m^{-1}(\xi+\frac kn)\|\right\}.
\eq
The submatrix $Q$ is an $m\times m(n-1)$ matrix with entries of norm $\frac {1}{mn}$. Thus, we have 
\begeq
\|Q\| \le m\sqrt{n-1}\|Q\|_{max} = \frac{\sqrt{n-1}}{n}.
\eq
Observing that the columns of $U_{k_0}$ are orthogonal, and we have $\|U_{k_0}^{-1}\| = mn$. Our estimate \eqref {est1} becomes, 
\begeq
\|A_\Omega^\ell(\xi) \| \le m \max\left\{ n, \max_{k\neq k_0}\|\mathcal A_m^{-1}(\xi+\frac kn)\|\right\}( 1 + m\sqrt{n-1}).
\eq

Taking the essential supremum over $\xi \in [0,\frac 1n ]$, and noting that for $k\neq k_0$,  $\xi+\frac k n\in J$ as in \eqref{Jdef},  this last equation can be estimated by
\begeq
\begin{array} {lll}
\|\mathbf A_\Omega^\dagger\|  &\le& \ssup\limits_{\xi\in [0,\frac 1 n]} \Big(m\max \left\{ n, \max_{k\neq k_0} \|\mathcal A_m^{-1}(\xi+\frac kn)\|\right\}\Big)( 1 + m\sqrt{n-1})\\
&\le&  m\max \left\{ n, \ssup\limits_{\eta\in J}\|\mathcal A_m^{-1}(\eta)\|\right\}( 1 + m\sqrt{n-1}).
\end {array}
\eq

 Since $\mathcal A_m(\eta)$ is invertible for all $\eta \in J$ and $J$ is a compact set, it follows that $\ssup_{\eta\in J} \|\mathcal A_m^{-1}(\eta)\|$ is finite, and   Theorem \ref{stabup} follows. 

To find the more explicit bound in Corollary \ref{cor1stabup}, we use the estimate for the norm of the inverse of a Vandermonde matrix \cite{G62}:
\begeq
\|\mathcal A_m^{-1}(\xi)\| \leq \sqrt{m}
\max_{0\leq i\leq m-1}\prod_{\begin{smallmatrix} j=0 \\ j\neq i \end{smallmatrix}}^{m-1}\frac{1+|\hat a(\frac{\xi+j}{m})|}{|\hat a(\frac{\xi+j}{m})-\hat a(\frac{\xi+i}{m})|}.
\eq

To prove Corollary \ref{cor2stabup}, we find a uniform lower bound for $|\hat a(\frac{\xi+j}{m} )-\hat a(\frac{\xi+i}{m})|$. Note that when $\xi \in J$, we have  $\frac{\xi+j}{m} \in \frac j m + \frac 1 m J$, 
$j=0,\dots,m-1$.  
Then for $\xi \in J$ and any $j=0,\dots,m-1$,  we have 
\[\frac{\xi+j}{m} \in \bigcup\limits_{j=1}^{m-1} \big\{\frac j m + \frac 1 m J\big\}\subset [\frac{1}{4mn}, \frac 12 - \frac{1}{4mn}] \cup [\frac 12 + \frac{1}{4mn}, 1 - \frac{1}{4mn}].\]
Thus, defining $M:=[\frac{1}{4mn}, \frac 12 - \frac{1}{4mn}]$, we have   that $\frac{\xi+j}{m} \in M \cup (M+\frac12)$ for any $j=0,\dots,m-1$.

Let $\gamma = \min\limits_{\xi \in M}|\hat{a}'(\xi)|>0$ where $\hat{a}'(\xi)$ denotes the first derivative of $\hat{a}(\xi)$. By the symmetry of $\hat a$, we also have $\gamma= \min\limits_{\xi \in M+\frac 1 2}|\hat{a}'(\xi)|$ . Without loss of generality, assume $\frac{\xi+j}{m} > \frac{\xi+i}{m}$. If the interval $[\frac{\xi+i}{m}, \frac{\xi+j}{m}]$ is contained in $M$ or in $M+ \frac 1 2$, the Mean Value Theorem gives
\begeq
\nonumber
\left|\hat a(\frac{\xi+j}{m})-\hat a(\frac{\xi+i}{m})\right| \geq \gamma \left|\frac{\xi+i}{m} - \frac{\xi+j}{m}\right|
\geq \gamma  \frac 1m. 
\eq 

If $\frac{\xi+i}{m} \in M$ and $\frac{\xi+j}{m} \in M+ \frac 1 2$, we exploit the symmetry of $\hat{a}$ and consider the interval between $1 - \frac{\xi+j}{m}$ and $\frac{\xi+i}{m}$, which is contained in $M$. Defining $l = m-i-j$ and using the Mean Value Theorem again, we have
\begeq\nonumber
\left|\hat a(1-\frac{\xi+j}{m})-\hat a(\frac{\xi+i}{m})\right| \geq \gamma \left|\frac lm - 2(\frac{\xi}{m})\right|\\ 
= \gamma \frac2m \left|\frac l2 - \xi\right|
\geq \gamma \frac{1}{2mn},
\eq
where the last inequality follows from the fact that $\l \in \ZZ$  and $\xi \in J$. This gives Corollary \ref{cor2stabup}. Notice that if $\hat a^\prime\in C(\TT)$ then $\gamma \rightarrow 0$ as $n \rightarrow \infty$, due to the fact that the minimum is taken over a larger interval getting closer to the zeros of $\hat{a}'$. 

\subsubsection {Proof of Theorem \ref {stablow}}Recall that $\| A^\dagger(\xi)\|$ is equal to the reciprocal of the smallest singular value of $ A(\xi)$, denoted $s_{min}( A(\xi))$. We choose an extra sampling set $\Omega$ according to Theorem \ref{AddSample}. We claim that\\
{ \it Claim 1} : There exists an interval $[0,r]\subset [0,\frac 1 {4n}]$, such that the smallest singular value of $s_{min}( A_\Omega(\xi))$ is bounded above on $ [0,r]$, by
\[ 0\le s_{min}( A_\Omega(\xi))\le \frac 1 {m\|\mathcal A_m^{-1}(\xi+\frac 1n)\|} < \infty, \quad \; \xi \in [0,r].\]

Using the claim, the theorem follows from
\begin{align}
\label {maininq}
m\|\mathcal A_m^{-1}(\frac 1n)\|& \le m\cdot \underset{\xi\in [0,r]}{\ssup} \,
 \|\mathcal A_m^{-1}(\xi+\frac 1n)\|\\ \nonumber 
&\le \underset{\xi\in [0,r]}{\ssup} \,  \frac 1 {s_{min}( A_\Omega(\xi))}\\ \nonumber 
  &\le \underset{\xi\in [0,\frac 1n]}{\ssup}\,  \frac 1 {s_{min}( A_\Omega(\xi))}\\ \nonumber
  &= \| \mathbf A_\Omega^\dagger\|. \nonumber
\end{align}

\noindent \emph {Proof of Claim 1}. 
We first show that $s^2_{min}( A_\Omega(\xi))$ is equal to the $mn$-th largest eigenvalue $\lambda_{mn}( A_\Omega(\xi)  A_\Omega^*(\xi))$ of $A_\Omega(\xi)  A_\Omega^*(\xi)$:
\begeq \label {astr}
A_\Omega(\xi)   A_\Omega^*(\xi) =  \frac1{m^2}
\left(
\begin{array}{ccc} 
\ast & \ast&\ast \\
\ast & \mathcal A_m(\xi)\mathcal A_m^*(\xi) & 0\\
\ast & 0 & D(\xi)D^*(\xi)
\end{array}\right),
\eq
where the matrices in the first row have $|\Omega|$ rows and $D(\xi)D^*(\xi)$ is the  block diagonal matrix with blocks $A_m(\xi+\frac {k}{n})\mathcal A_m^*(\xi+\frac {k}{n}), k \neq 0$, as entries. The rank of the $(mn + |\Omega|) \times (mn + |\Omega|) $ matrix $ A_\Omega(\xi)  A_\Omega^*(\xi)$ is equal to the rank of $ A_\Omega(\xi)$, which is $mn$. Thus, the smallest positive  eigenvalue of $ A_\Omega(\xi)  A_\Omega^*(\xi)$ 
is the  $mn$-th largest eigenvalue $\lambda_{mn}( A_\Omega(\xi)  A_\Omega^*(\xi))$, and it is equal to $s^2_{min}( A_\Omega(\xi))$. 
Thus, to estimate $s^2_{min}( A_\Omega(\xi))$ from above, we need to estimate $\lambda_{mn}( A_\Omega(\xi)  A_\Omega^*(\xi))$. 

In turn, the  $mn$-th largest eigenvalue $\lambda_{mn}( A_\Omega(\xi)  A_\Omega^*(\xi))$ can be estimated above using the eigenvalues of the $mn\times mn$ principal submatrix  $B(\xi)$ 
\begeq
B(\xi)=\left(\begin{smallmatrix}  \mathcal A_m(\xi+\frac {k_0}{n})\mathcal A_m^*(\xi+\frac {k_0}{n})   
& 0\\
 0 & D(\xi)D^*(\xi)\end{smallmatrix}\right),
\eq 
via the Cauchy Interlacing Theorem \cite{FP57}:
\begeq
\label{untest}
s^2_{\min}( A_\Omega(\xi)) = \lambda_{mn}( A_\Omega(\xi)A_\Omega^*(\xi)) \le \frac{1}{m^2}\lambda_{mn-|\Omega|}(B(\xi)),
\eq
where we use $\lambda_{j}(M)$ to denotes the $j$-th largest eigenvalue of the a matrix $M$ counting the multiplicity. 

We chose $\Omega$ to be a minimal extra sampling set so that $|\Omega| = \frac{m-1}{2}.$ Observing that $B(\xi)$ is block diagonal so that the eigenvalues of $B(\xi)$ are the eigenvalues its $\mathcal A_m(\xi+\frac kn)\mathcal A_m^*(\xi+\frac kn)$, and using a continuity argument below we show that there exists $r$ with $0<r<\frac 1 {4n}$ such that for all $\xi\in[0,r]$
\begin{align}
\label {test1}
 \lambda_{mn-\frac {m-1} 2}(B(\xi))
&=\min_{k\neq 0} \{\lambda_m\Big(\mathcal A_m(\xi+\frac kn)\mathcal A^*_m(\xi+\frac kn)\Big)\} \\
&\leq\lambda_m\Big(\mathcal A_m(\xi+\frac 1n)\mathcal A^*_m(\xi+\frac 1n)\Big) \nonumber \\
&= \frac  1 {\|\mathcal A_m^{-1}(\xi+\frac 1n)\|^2}. \nonumber
\end{align}
In the last equality above, we  used the relation between the minimum singular values of a matrix $M$ and the norm of its inverse:   $s^{-1}_{min}( M)  =\|M^{-1}\|$.  Claim 1 then follows from \eqref {untest} and \eqref {test1}.  
 
We now use the continuity argument to prove \eqref {test1}. 
Let $$\alpha := \inf_{[0, \frac {1}{4n}]} \lambda_{m(n-1)}\Big(D(\xi)D^*(\xi)\Big) > 0.$$
Since  $\lambda_{m-j}\Big(\mathcal A_m(0)\mathcal A_m^*(0)\Big) = 0$ for $j=0,\dots,(\frac {m-1} 2-1)$, continuity in $\xi$ implies that  there exists $r$ with $0<r<\frac {1}{4n}$ such that 
$$\lambda_{m-j}\Big(\mathcal A_m(\xi)\mathcal A_m^*(\xi)\Big) < \alpha \,\, \text{for} \, \xi \in [0, r].$$ 
Thus, when $\xi \in [0, r]$, the smallest $\frac {m-1} 2$ eigenvalues of $B(\xi)$ are precisely the smallest $\frac {m-1} 2$ eigenvalues of $\mathcal A_m(\xi)\mathcal A^*_m(\xi)$, i.e.,  $\lambda_{mn-j}\Big(B(\xi)\Big) = \lambda_{m-j}\Big(\mathcal A_m(\xi)\mathcal A_m^*(\xi)\Big)$ for $j=0,\dots,(\frac {m-1} 2-1)$ and

\begin{align*}
\lambda_{mn-\frac {m-1}2}\Big(B(\xi)\Big) &= \min \left\{ \lambda_{m-\frac {m-1}2}\Big(\mathcal A_m(\xi)\mathcal A_m^*(\xi)\Big), \lambda_{m(n-1)} \Big(D(\xi)D^*(\xi)\Big)\right\}\\
&\le \lambda_{m(n-1)} \Big(D(\xi)D^*(\xi)\Big)\\
&= \min_{k\neq 0} \{\lambda_m\Big(\mathcal A_m(\xi+\frac kn)\mathcal A^*_m(\xi+\frac kn)\Big)\}\\
&\le \lambda_m\Big(\mathcal A_m(\xi+\frac 1n)\mathcal A^*_m(\xi+\frac 1n)\Big)\\
&= \frac  1 {\|\mathcal A_m^{-1}(\xi+\frac 1n)\|^2},
\end{align*}
which is \eqref {test1}.

\section{Acknowledgments}
We would like to thank Rosie the cat for engaging Penelope the toddler in many chase games, thereby leaving us time to write this manuscript.

\bibliographystyle{siam}
\bibliography{../refs}

\end{document}